\documentclass[fleqn,usenatbib]{mnras}

\usepackage{newtxtext,newtxmath}

\usepackage[T1]{fontenc}

\DeclareRobustCommand{\VAN}[3]{#2}
\let\VANthebibliography\thebibliography
\def\thebibliography{\DeclareRobustCommand{\VAN}[3]{##3}\VANthebibliography}

\usepackage{graphicx}	
\usepackage{amsmath}	
\usepackage{siunitx}
\usepackage{makecell}
\usepackage{tabularx}
\usepackage{cleveref}
\newcolumntype{Y}{>{\centering\arraybackslash}X}

\DeclareSIUnit\year{yr}

\usepackage{xcolor}
\usepackage{soul}

\newcommand{\change}[1]{#1}
\newcommand{\changetwo}[1]{#1}

\title[Direct $N$-body simulations of NGC 6397]{Direct N-body simulations of NGC 6397 and its tidal tails}

\author[A. D. Arnold et al.]{
Anthony D. Arnold,$^{1}$\thanks{E-mail: anthony.arnold@uqconnect.edu.au (ADA)}
and Holger Baumgardt$^{1}$
\\
$^{1}$School of Mathematics and Physics,The University of Queensland, St. Lucia 4072, Queensland, Australia
}

\date{Accepted XXX. Received YYY; in original form ZZZ}

\pubyear{2024}

\begin{document}
\label{firstpage}
\pagerange{\pageref{firstpage}--\pageref{lastpage}}
\maketitle

\begin{abstract}
We have performed a series of direct $N$-body simulations that study the evolution of the Galactic globular cluster NGC~6397 under the combined influence of two-body relaxation, stellar evolution and the Milky Way's tidal field. Our simulations follow the evolution of the cluster over the last several Gyr up to its present-day position in the Milky Way in order to allow us to derive present-day cluster parameters and the distribution of its extra-tidal stars. We have also determined a new density profile of NGC 6397 by selecting stars from Gaia DR3 using Gaia DR3 proper motions, parallaxes and photometry to discriminate cluster members from non-members. This allows us to derive the surface density profile of NGC~6397 and the location of its tidal tails up to \ang{10} of the cluster centre, well beyond the tidal radius of NGC~6397. Our results show that the current state of NGC 6397 in terms of surface density, velocity dispersion profile and stellar mass function can be matched by a cluster model evolving from a standard initial mass function and does not require an additional central cluster of dark remnants. We also find good agreement in the location and absolute number of the extra-tidal stars between our simulations and the observations, making it unlikely that NGC 6397 is surrounded by a dark matter halo.
\end{abstract}

\begin{keywords}
galaxy dynamics – (Galaxy:) globular clusters: individual: NGC 6397 – Milky
Way – methods: orbital integrations – $N$-body simulation
\end{keywords}



\section{Introduction}

Globular clusters continue to be important stellar laboratories and contain exotic astrophysical objects such as black holes~\citep{2018MNRAS.475L..15G, Chomiuk_2013} and millisecond pulsars~\citep{2013IAUS..291..243F,2008IAUS..246..291R}. Some globular clusters also show evidence of containing intermediate-mass black holes~\citep{Häberle2024}, an elusive class of stellar exotica which bridge the mass gap between stellar-mass and supermassive black holes. Furthermore, far from being simple isotropic systems in virial equilibrium, some globular clusters have been found to undergo tidal disruption by the tidal field of the Milky Way, which gradually strips away stars from the outer region, leading to extended stellar halos~\citep{2017MNRAS.471L..31P,2019MNRAS.485.4906D} and tidal tails~\citep{2021ApJ...914..123I}. Since globular clusters might have formed inside small dark matter halos~\citep{Carlberg_2021, Carlberg_2022} which would protect stars from tidal stripping, it has been suggested to use the presence or absence of tidal tails  to test for the presence of dark matter in globular clusters~\citep{2021MNRAS.507.1814B}. Dark matter has also been put forward as a possible explanation for the velocity dispersion profile of NGC 3201, which shows a decrease in its outer parts which is slower than expected based on the visible stellar content of the cluster alone~\citep{2021MNRAS.507.1814B,10.1093/mnras/stab306}. 

It is therefore important to accurately model globular clusters in order to better understand their evolution and the role that dark matter could play in them. In particular, an explanation for the lack of visible tidal tails where simulations expect them to exist could provide clues to the evolution of the host galaxy and the nature of dark matter. Recently, \citet{2023MNRAS.519..192W} investigated four globular clusters using gravitational $N$-body simulations and new observational data and found a good agreement between both, arguing against significant dark matter in the outer parts of the four studied clusters.

In this work we consider NGC 6397, which is a well studied globular cluster and the second closest globular cluster to the Sun~\citep{2021MNRAS.505.5957B}. NGC 6397 is a core-collapsed, low-mass, low-velocity galactic globular cluster which has been predicted to contain a small diffuse inner dark component~\citep{2016A&A...588A.149K,Vitral&Mamon,2022MNRAS.514..806V}. ~\citet{2021MNRAS.507.1814B} simulated a cluster similar to NGC 6397 with the GOTHIC~\citep{Miki_2017} $N$-body gravitational code, a GPU-based tree code with hierarchical time steps. They found that NGC 6397 should have formed a strong extended tail in less than \SI{4}{\giga yr} which they could not find in Gaia DR3 data. \citet{2021MNRAS.507.1814B} stress that ``a DM minihalo could be responsible for both the inner spherical shape and the absence of tidal tails in this cluster, as the stellar part of a GC embedded in DM would be more resilient to tidal disruption.''

In this paper we use direct $N$-body simulations to predict the surface density and velocity dispersion profile of NGC 6397 and compare the resulting distribution with observational data from Gaia DR3. In the inner regions of NGC~6397, where the Gaia data is not sufficient, we supplement the Gaia velocity data with the HST based velocity dispersion profile from~\citet{2022ApJ...934..150L} as well as stellar radial velocities from~\citet{2018MNRAS.478.1520B}. In our paper we aim to perform simulations of NGC 6397 with the exact number of real stars, in order to avoid having to do any corrections due to a different number of cluster stars. We furthermore set up our clusters on the exact same orbit as NGC 6397 and stop the simulations once the current position has been reached. Our simulations should therefore be the most realistic simulations done for NGC~6397.

Our paper is organised as follows: Section~\ref{sec:observations} discusses how we obtained the observational data and selected member stars for NGC~6397 and presents the resulting surface density and velocity dispersion profiles of the cluster. Section~\ref{sec:nbody} describes the code used and the setup of the simulations and section~\ref{sec:results} compares the observed and simulated velocity dispersion and surface density profiles of the cluster. It also discusses the cluster disruption and the formation of tidal arms over time. In section~\ref{sec:discussion} we discuss the differences between this work and previous efforts to simulate NGC 6397 and present our conclusions.

\section{Observational Data}
\label{sec:observations}

\subsection{Surface Density Profile}
\label{subsec:sb}

In this section we describe the observational data that we have used to derive the surface density profile of NGC 6397. There is a wealth of available data for this cluster, and we make use of the photometry from ~\citet{2007AJ....133.1658S} and ~\citet{2019MNRAS.485.3042S} to determine the surface density profile in the inner cluster parts. We also downloaded from the Gaia database~\citep{2023A&A...674A...1G} stars in a rectangular field of size \ang{30} $\times$ \ang{12} around the position
$(\alpha / \delta) = (\ang{265.17538}/\ang{-53.67434})$, corresponding to the centre of NGC 6397 as found by \citet{2010AJ....140.1830G}. After removing stars without proper motion information, this search results in $5.53\times 10^7$ stars. \citet{2023A&A...669A..55C} found that the completeness of the Gaia data is not uniform across the sky for the faintest stars, but it is generally better than 90\% for stars with $G<19.8$. For this reason, we \change{initially selected} only stars with $G<19.5$ and a total proper motion error $<1$ mas/yr. 

After downloading the data, we first selected member stars based on their proper motions. Considering the size of our Gaia field, the proper motions of the cluster members are changing across the field due to the change in perspective and the overall motion of the stars along the tidal arms. Hence assuming a single proper motion of all cluster and tail stars is not sufficient. We therefore used the simulations from Section~\ref{sec:nbody} to fit two 2$^{nd}$ order polynomials to the proper motion of stars in the tails to the left and right of the cluster. Since the tails in the vicinity of NGC~6397 are mainly elongated in right ascension, we used polynomials as a function of right ascension only. Using these polynomials, we then applied the following test to the Gaia data:%
\begin{equation}\label{eq:pmcut}
 \chi_{pm}^2 =  \frac{(\mu_\alpha \cos\delta_i - \mu_\alpha \cos\delta)^2}{\epsilon_{RA,i}^2+\epsilon_{RA}^2} + \frac{(\mu_{\delta_i} - \mu_{\delta})^2}{\epsilon_{DEC,i}^2+\epsilon_{DEC}^2} \text{,}
\end{equation}%
where $\mu_\alpha \cos\delta_i$ and $\mu_{\delta_i}$ are the proper motion of individual stars in right ascension and declination respectively, and $\mu_\alpha \cos\delta$ and $\mu_{\delta}$ are the average proper motion of stars in our simulations of NGC~6397 at the right ascension of the observed stars. $\epsilon_{RA,i}$ and $\epsilon_{DEC,i}$ are the individual proper motion errors of each star in right ascension and declination. The terms $\epsilon_{RA}$ and $\epsilon_{DEC}$ reflect the fact that in our simulations the stars in the tails show spreads in proper motion at any given position due to the stars being on different orbits and were set equal to $\epsilon_{RA} = 0.20$ mas/yr and $\epsilon_{RA} = 0.50$ mas/yr. We only consider stars with $\chi_{pm}^2<2$ as cluster members. We varied the above values but found that our results for the density profile do not strongly depend on the exact choice of these values.

We next performed a parallax selection on the potential member stars, taking only stars with%
\begin{equation}
    \label{parallax}
    \chi_{par}^2 = \frac{(\pi_i-\pi_{Cl})^2}{\epsilon_{\pi}^2+0.05 \mbox{mas}^2} < 3 \text{,}
\end{equation}%
as potential cluster members. In Equation~\eqref{parallax} $\pi_i$ and $\epsilon_{\pi}$ are the parallax and parallax uncertainty of each star and we used the cluster distance from \citet{2021MNRAS.505.5957B} to get $\pi_{Cl}=1/D_{Cl}=0.403$ mas. We added an additional uncertainty of 0.05 mas to the denominator of eq.~2 to account for the finite width of the tails and the cluster.

\begin{figure}
 \center
    \includegraphics[width=\columnwidth]{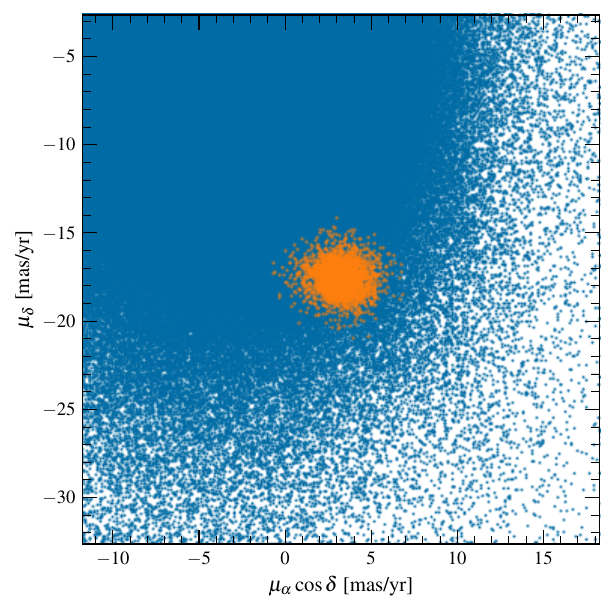}\\
    \includegraphics[width=\columnwidth]{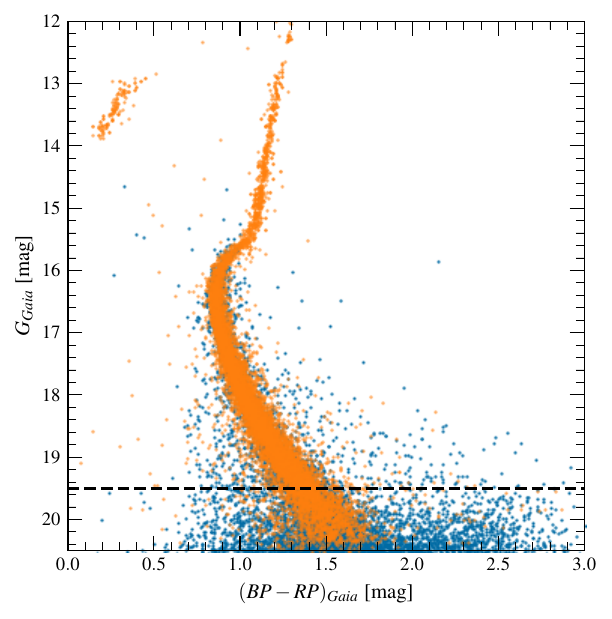}
    \caption{\textit{Upper panel:} Proper motion vector point diagram of all Gaia DR3 stars within \ang{3} of the cluster centre. Background stars are shown in blue and stars that pass our proper motion and parallax membership criteria are shown in orange. \textit{Lower panel:} Gaia $BP-RP$ vs. $G$ colour-magnitude diagram of the stars that pass parallax and proper motion criteria. Background stars are shown in blue and stars which pass all of our membership criteria are shown in orange. The black dashed line marks the limit of $G=19.5$ mag. Only stars brighter than this limit are used to calculate the surface density profile.}
    \label{fig:pop}
\end{figure}

 We finally fitted a PARSEC isochrone \citep{2012MNRAS.427..127B} to the colour-magnitude diagram (CMD) of the stars that pass the proper motion and parallax tests,
 selecting stars that deviated in BP-RP color by less than 0.15 mag or less than 1.5 times their photometric error from the isochrone. In addition, we selected  a rectangular region in the upper left corner of the CMD to select the horizontal branch stars of NGC~6397. Figure~\ref{fig:pop} shows the proper motion diagram of stars inside \ang{3} of the cluster centre as well as the final 
 colour-magnitude diagram of stars that passed all membership criteria. We ended up with a final population of 8,330 potential members.%

In order to calculate the surface density profile of NGC~6397 from the Gaia DR3 members \change{in terms of numbers or stars per arcsec$^2$}, we considered only bright stars with $G < 17.8$ to reduce incompleteness issues in the inner parts of the cluster. As will be shown in Section~\ref{sec:results}, the density of Gaia member stars decreases strongly for right ascension values less than \ang{260}, while no such decrease is seen in our simulations. This can probably be explained by the low galactic latitude of the tidal tail of NGC~6397 in this direction which leads to a strong extinction of the tail stars. We therefore discarded all data with RA$<\ang{260}$ when calculating the surface density profile outside the tidal radius.  We binned the bright stars according to their radial distance from the cluster center, calculated the number of stars in each bin and divided this number by the area of the annulus that is covered by our Gaia field.

\begin{figure}
    \centering
    \includegraphics[width=\columnwidth]{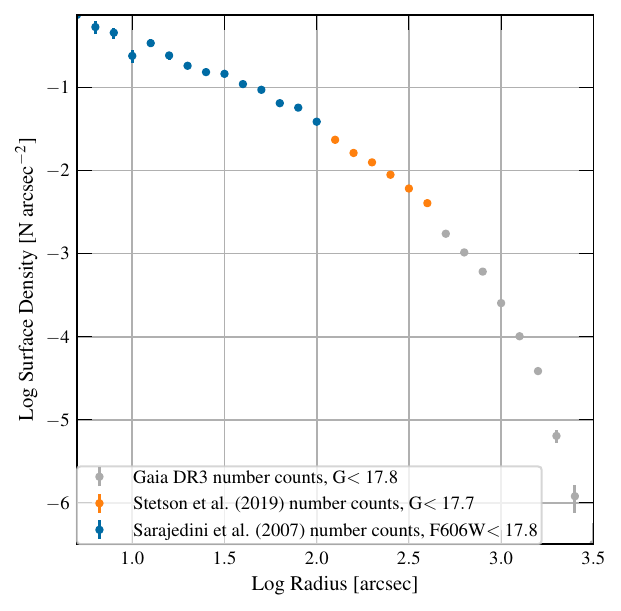}
    \caption{The combined \change{background subtracted} surface density profile from Gaia DR3 \change{for $G < 17.8$} (grey points with error bars) for the outer radii plus the surface density profiles derived by~\citet{2007AJ....133.1658S} \change{for F606W$<17.8$} (blue points) at the innermost radii and~\citet{2019MNRAS.485.3042S} \change{for $G<17.7$} (orange points) for the intermediate radii.}
    \label{fig:sb}
\end{figure}

\change{Since the Gaia catalogue is incomplete in areas of high surface density, we discarded the Gaia data inside $450\arcsec$ and supplemented it with Hubble Space Telescope photometry from}~\citet{2007AJ....133.1658S} \change{inside 100 arcsec. To bridge the gap between $100\arcsec$ and $450\arcsec$, we used the ground-based photometry from}~\citet{2019MNRAS.485.3042S}. The combined surface density profile that we have derived from all three data sets is shown in Figure~\ref{fig:sb}. \change{Because we want to compare our simulations to the observed profile, we needed to select a similar mass cutoff in the simulated and observed data. As discussed in Section}~\ref{subsec:sd-vd}, \change{we use a mass cutoff of $0.7 M_\odot$ to select bright stars from our simulations corresponding to a $G<17.8$ brightness cutoff in the Gaia data. We similarly chose brightness cutoffs of F606W$<17.8$ and $V<17.7$ for the data from}~\citet{2007AJ....133.1658S} and~\citet{2019MNRAS.485.3042S}, respectively.

We determined the background density from the potential member stars found at declinations $<$-\ang{56} and $>$-\ang{50} since our simulations show that only few stars in the tidal tails should be located in this region. Since the density of background stars in this area increases towards large right ascension values, we calculated the surface density of background stars as a function of right ascension in steps of \ang{4}. When calculating the total surface density, we also summed up the expected background density, considering for each central distance how much area we cover at each right ascension value. The surface density profile \change{with the background density subtracted} is shown in Fig.~\ref{fig:sb} and given in Table~\ref{tab:sb}.%

\subsection{Kinematic Data}
\label{subsec:vd}

\begin{figure}
    \centering
    \includegraphics[width=\columnwidth]{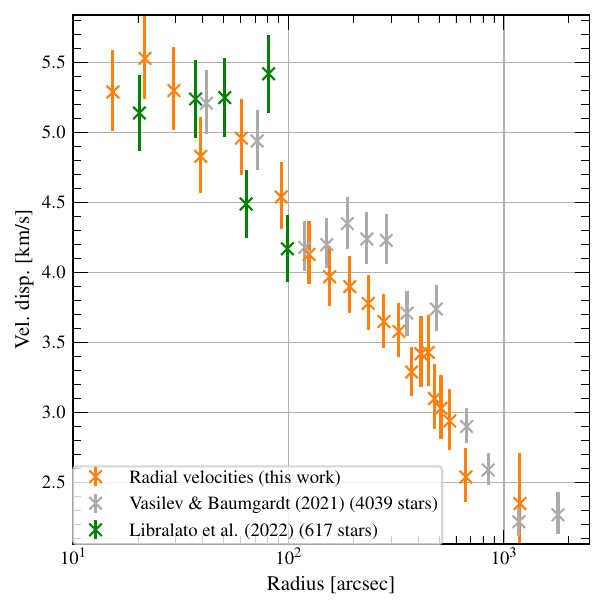}
    \caption{Velocity dispersion profile of NGC 6397 used in this work. The radial velocity dispersion profile of NGC 6397 calculated in this work is shown in orange, while the Gaia proper motion profile taken from \citet{2021MNRAS.505.5978V} is shown in grey. The HST based proper motion velocity dispersion profile from \citet{2022ApJ...934..150L} is shown in green.}
    \label{fig:vd}
\end{figure}

In addition to the surface density profile, we also compiled kinematic data (radial velocities and proper motions) from multiple sources to compare this data with our simulations. The data is shown in Figure~\ref{fig:vd} and consists of radial velocities from~\citet{2018MNRAS.478.1520B} plus the updates reported in \citet{2023MNRAS.521.3991B}. %
In total we found 4,039 cluster stars with measured radial velocities from which we calculated
the velocity dispersion profile using the method described in~\citet{2018MNRAS.478.1520B}. We further used the Gaia proper motion velocity dispersion profile from~\citet{2021MNRAS.505.5978V} in the outer cluster parts and the HST based proper motion velocity dispersion profile from~\citet{2022ApJ...934..150L} in the inner cluster parts. We converted both proper motion velocity dispersion profiles into physical velocity dispersions using $d=2480$ pc as cluster distance~\citep{2021MNRAS.505.5957B}.

\change{We note the relative large scatter of the measured velocity dispersion profile in Figure}~\ref{fig:vd} \change{from}~\citet{2022ApJ...934..150L} \change{outside $60$\arcsec. Their final three data points describe large changes in velocity dispersion which are unlikely to be physical. These data points could either be extreme statistical outliers or perhaps be influenced by other effects. Importantly, as we describe in Section}~\ref{subsec:sd-vd}\change{, our $N$-body velocity dispersion profile fits in between these extreme points in Figure}~\ref{fig:vd}.

\section{N-body Simulations}
\label{sec:nbody}
We performed gravitational $N$-body simulations of NGC~6397 using a significantly modified version of the NBODY6 code~\citep{10.1111/j.1365-2966.2012.21227.x}, to prepare comparisons of the velocity dispersion and surface density profiles. Our code, \change{based on} NBODY6+P3T~\citep{2022MNRAS.509.2075A}, uses a \change{single} leapfrog scheme for the particle integration \change{instead of the dual regular and irregular Hermite integrators from NBODY6}, and the GPU-based tree code from~\citet{2012ASPC..453..325B} for force calculations. \change{Close encounters between stars are handled with softening instead of regularisation.} It retains all other parts (data input/output, stellar evolution, etc.) from NBODY6. To simulate the effects of the Milky Way's tidal field, we included the Milky Way mass models of~\citet{2015ApJS..216...29B} and~\citet{Irrgang2013} into our code. Both models were included as external tidal fields in addition to the existing models in NBODY6. We also modified the stellar evolution code from~\citet{2000MNRAS.315..543H, 2002MNRAS.329..897H} to include a run-up phase for stellar evolution. During this run-up phase stellar evolution is applied to all stars up to a user specified time without integrating their orbits. When the full $N$-body gravitational simulation starts, the stellar evolution continues from the end of the run-up phase. The length of the run-up phase was chosen as%
\begin{equation}
    T_{\text{run}} = T_C - T_{\text{Sim}},
\end{equation}%
where $T_{\text{Sim}}$ is the $N$-body integration time and $T_C$ is the age of NGC~6397,
assumed to be \SI{13}{\giga yr}~\citep{2013ApJ...775..134V}. This way we can include stellar evolution in our simulations without having to simulate the evolution of the clusters over a full Hubble time.

In order to initialise our simulations, we followed a similar procedure to the one used by~\citet{2023MNRAS.519..192W} to investigate tidal tails in other globular clusters. We took the mean proper motion and distance of NGC 6397 from~\citet{2019MNRAS.482.5138B} and ~\citet{2021MNRAS.505.5957B}, respectively, and used a fourth-order Runge-Kutta scheme to integrate the cluster orbit in the chosen Milky Way potential backward in time for a given time $T_{\text{Sim}}$. We then initialised $N$-body models of different masses and sizes, using the initial mass functions from~\citet{2023MNRAS.521.3991B}, based on the grid of $N$-body models described in~\citet{2018MNRAS.478.1520B} and then performed a full $N$-body simulation of the cluster forward in time up to the present time. \change{The models in} \citet{2018MNRAS.478.1520B} \change{are scaled to fit globular clusters. Assuming two-body relaxation is the main driver of cluster evolution, the models are moved in a mass vs. radius plane along lines of constant relaxation time. That way clusters with different masses and sizes can be modeled.} The simulations were stopped when the simulated cluster reached its minimum distance from the present-day position of NGC 6397. The deviations of the simulated cluster and the true position of NGC~6397 are given in column 9 of Table~\ref{table:runs}. They are generally smaller than 3 pc, which is well within the half-mass radius of the cluster and significantly less than the distance of the cluster from the Galactic centre. We therefore expect that tidal tails found in our simulations should be very close to the locations of the true tails of NGC~6397.

Since we want to investigate the formation of tidal tails in this paper, we kept all stars, including stars unbound to the simulated cluster, in the simulation. The final positions and velocities of all stars in each simulation were projected onto the sky and we then derived the surface density and velocity dispersion profiles of the clusters. We also determined the bound mass and the number of bound stars left in the clusters at the end of each simulation to derive an initial cluster mass and total lifetime of NGC~6397 \change{, given in Section}~\ref{subsec:massloss}. In order to compare with the surface density and velocity dispersion profiles of NGC~6397, we selected the run which best fits the observational data. We ran simulations for either 4 or 8 Gyr, but found little difference in the final profiles. Some runs were also performed without stellar evolution since we don't expect stellar evolution to have a large influence on our results as the amount of mass lost due to stellar evolution is very small over the course of our simulations. All runs were performed using the \citet{2015ApJS..216...29B} potential until a best fit was found, and then the best-fitting simulation was re-run using the \citet{Irrgang2013} potential for comparison. The ultimate best fit cluster was obtained from a \SI{4}{\giga yr} run, with stellar evolution enabled and in the ~\citet{2015ApJS..216...29B} potential. The simulation parameters for the different runs are given in Table~\ref{table:runs}.

\begin{table*}
\centering
\caption{\change{Parameters of the simulations done in this paper at the start and end of the run.} From left to right, the columns indicate: simulation number; the number of cluster stars; whether or not stellar evolution was enabled; the initial cluster mass; the initial half-mass radius; the ratio of final bound mass to initial mass; the ratio of total mass lost due to stellar evolution to the initial mass; the final half-mass \change{radius}; the 3D distance of the final cluster location from the true location of NGC 6397; the simulation time of the runs. All runs shown were done in the \citet{2015ApJS..216...29B} potential, with the exception of the run marked with (*) which used the \citet{Irrgang2013} potential. The runs which resulted in the best fits to the observational data are shown in bold. Run 8 started with the same initial condition as run 4, except that it was done in the \citet{Irrgang2013} potential.}
\label{table:runs}

\begin{tabularx}{\textwidth}{YYYYYYYYYY}
\hline
\thead{Run} &  
\thead{$N$} & 
\thead{SEV} & 
\thead{$M_{\text{ini}}$ \\ $\left[M_{\odot}\right]$} & 
\thead{$R_{H,\text{ini}}$ \\ $\left[ \text{pc} \right]$} & 
\thead{$M_{\text{bound,f}}/$ \\ $M_{\text{ini}}$} & 
\thead{$\Delta M_{\text{loss}}/$\\ $M_{\text{ini}}$} &
\thead{$R_{H,\text{fin}}$ \\ $\left[ \text{pc} \right]$} & 
\thead{$\Delta x$\\ $\left[ \text{pc} \right]$}  &
\thead{$T_{\text{Sim}}$ \\ $\left[ \text{Gyr} \right]$}\\
\hline
1 &     144765 & \checkmark &	85202.2 &	3.90 &	0.78 & 0.02 & 4.30 & 2.58 & 4\\
2 &     170000 & &	96285.1 &	3.54 &	0.82 & &	3.74 & 1.88 & 4\\
3 &     180000 & &	101848.3 &	4.00 &	0.57 &	& 7.20 & 1.93 & 8\\
\textbf{4} &  \textbf{200000} &	\checkmark & \textbf{117681.3} &	\textbf{4.29} &	\textbf{0.79} &	\textbf{0.02} & \textbf{4.70} & \textbf{2.70} & \textbf{4}\\
5 &     220000 & &	124481.3 &	4.00 &	0.56 & &	7.86 & 1.53 & 8\\
6a &     220000 & &	125489.1 &	4.00 &	0.82 & &	4.18 & 1.09  &	4\\
6b &     220000 & &	125489.1 &	4.00 &	0.76 & &	4.40 & 2.35  &	8\\
7a &     273985 & &	128220.6 &	5.57 &	0.75 &	& 5.87 & 2.62 & 4\\
7b &     273985 & &	128220.6 &	5.57 &	0.67 &	& 6.34 & 3.08 & 8\\
\textbf{8*} &    \textbf{200000} &	\checkmark & \textbf{117681.3} &	\textbf{4.29} &	\textbf{0.78} &	\textbf{0.02} & \textbf{4.68} & \textbf{1.50} & \textbf{4}\\
\hline
\end{tabularx}
\end{table*}

\section{Results}
\label{sec:results}

\subsection{Surface density and velocity dispersion profile of NGC 6397}
\label{subsec:sd-vd}

\begin{figure}
    \centering
    \includegraphics[width=\columnwidth]{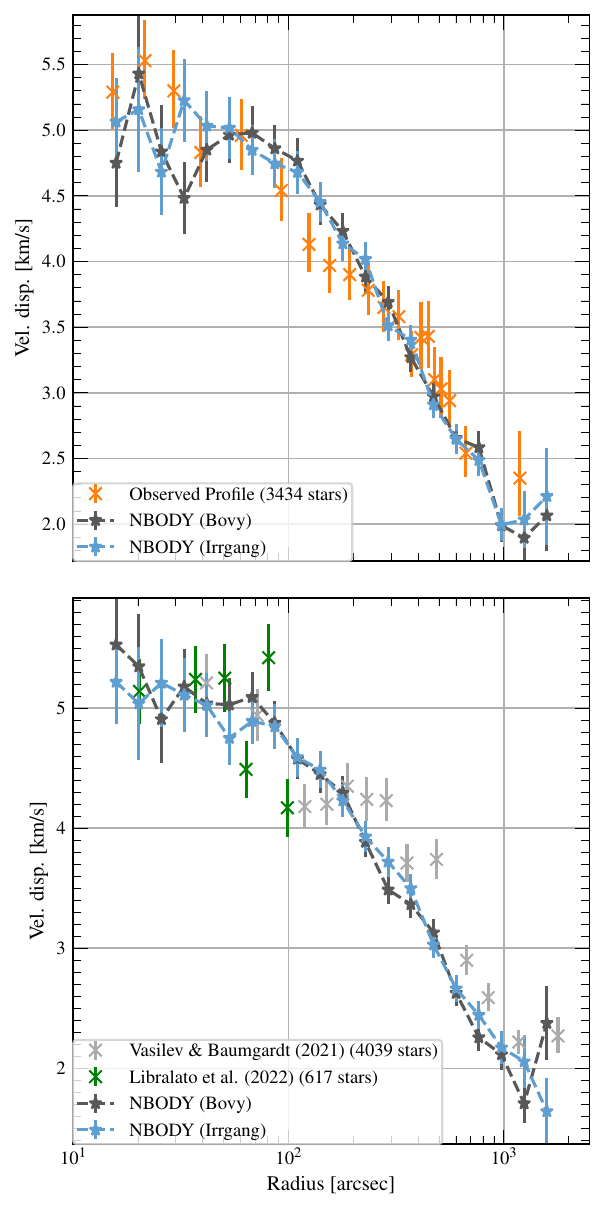}
    \caption{\textit{Upper panel}: Measured radial velocity dispersion profile of NGC 6397 against the $N$-body profile from runs \textbf{4} and run \textbf{8} for the ~\citet{2015ApJS..216...29B} and \citet{Irrgang2013} potentials respectively. \textit{Lower panel}: The same as the upper panel for the proper motion velocity dispersion profile. The observed profile for NGC 6397 is a combination of the HST based profile from~\citet{2022ApJ...934..150L} in the inner cluster parts and the Gaia DR3 based one in the outer cluster parts.}
    \label{fig:veldisp-nb}
\end{figure}

We calculated the velocity dispersion profile for both radial velocities and proper motions of the simulated clusters at the end of each run. We selected bound giant and main sequence stars above \change{$0.7 M_{\odot}$} in the direction of the cluster and calculated the mean radial velocity from them. \change{The mass cutoff was chosen to correspond to the brightness cutoff of $G < 17.8$ in the Gaia data from Section}~\ref{sec:observations}. We removed outliers which differed from the mean velocity by more than \SI{25}{\km\per\s}. We then binned the selected members in radius and calculated the velocity dispersion%
\begin{equation}
    \sigma = \sqrt{\frac{1}{n} \sum_{i=1}^{n} (V_i - \overline{V})^2}\text{,}
\end{equation}%
where $n$ is the number of stars in each bin, $V_i$ is the velocity of an individual star, and $\overline{V}$ is the mean velocity of the whole cluster. Figure~\ref{fig:veldisp-nb} shows the fit of the observed velocity dispersion profile for the best-fitting $N$-body simulations in 
the \citet{2015ApJS..216...29B} and \citet{Irrgang2013} potentials. It can be seen that the velocity dispersion profile of the simulations fit the observations well \change{in terms of radial velocity with the exception of a slight overprediction of the velocity dispersion around 100\arcsec. We also get a good fit of the proper motion velocity dispersion profile except for the outermost data points of}~\citet{2022ApJ...934..150L} \change{which was already pointed out in Section}~\ref{subsec:vd}. Run \textbf{8} gives a very similar velocity dispersion profile to run \textbf{4}, showing that the choice of Galactic potential has only a small influence on our results.%

\begin{figure}
    \centering
    \includegraphics[width=\columnwidth]{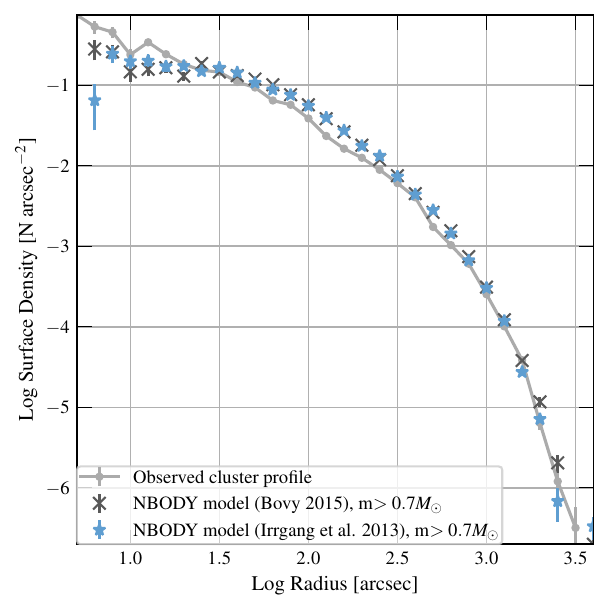}
    \caption{The surface density profile of NGC 6397 compared against the surface density profiles of the simulated clusters at the end of the simulations. The observed cluster profile is the same as in Figure~\ref{fig:sb}. }
    \label{fig:sb-nb}
\end{figure}

To further compare the simulated clusters with the observational data of NGC 6397, we produced a surface density profile of the clusters at the end of each run, and show the result in Figure~\ref{fig:sb-nb}. We calculated the surface density of the simulated clusters using the giant and main sequence stars above $0.7 M_\odot$ and used the same method as described in subsection~\ref{subsec:sb} for the real cluster. It can be seen that both runs \textbf{4} and \textbf{8} show a good fit of the surface density profile of NGC 6397. The cluster in the galactic potential given by~\citet{2015ApJS..216...29B} \change{contained 10,470 main sequence and giant stars with $>0.7M_\odot$ within a projected distance of $3350\arcsec$ from the centre, compared to 9,965 member stars in NGC~6397 in the combined Gaia, Stetson and HST catalogues within this radius}. In addition, the break in the observed surface density profile at radii of about 4000 arcsec as well as the overall surface density of stars outside the tidal radius is well reproduced in our simulations.%

\subsection{Central dark component}

We next investigated the presence of a central dark component in NGC 6397. \change{While investigating the presence of dark matter in globular clusters,}~\citet{1996IAUS..174..303H} \change{suggested that approximately $50\%$ of the mass in NGC 6397 is unobserved and is likely made up of mostly white dwarfs.} ~\citet{2009MNRAS.397L..46H} \change{investigated the composition of the core of NGC 6397 using $N$-body simulations and also found evidence that the core consists of around 50\% white dwarfs.} \citet{Vitral&Mamon} investigated the presence of an IMBH or cluster of unresolved objects in NGC 6397, finding evidence for a diffuse dark inner cluster of unresolved objects with a total mass of $1000$ to $2000 M_\odot$ and suggesting that this subcluster should be mostly made up of stellar mass black holes.  ~\citet{2022MNRAS.514..806V} further constrained the mass of the subcluster to $807^{+123}_{-323} M_\odot$ and its radius to $0.05$ pc using their MAMPOSSt-PM code to fit the cluster~\citep{2013MNRAS.429.3079M}. They furthermore suggested that the inner part of NGC~6397 is made up of hundreds of white dwarfs. \citet{2022MNRAS.514..806V} also used the Cluster Monte Carlo (CMC) code of~\citet{2022ApJS..258...22R} to fit NGC~6397, which resulted in a much lower density central cluster with a mass of about $800 M_\odot$ and a radius around $0.26$ pc. 

\begin{figure}
    \centering
    \includegraphics[width=\columnwidth]{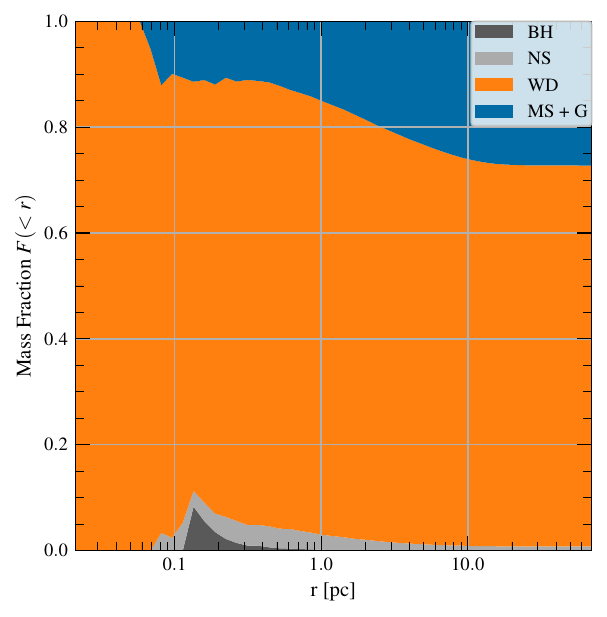}
    \caption{The fraction $F$ of the total cluster mass up to distance $r$ from the cluster centre for different stellar categories in run \textbf{4}. The proportion of mass from main sequence and giant stars is shown in blue, that of the white dwarfs in orange, that of neutron stars in grey, and that of black holes in black.}
    \label{fig:remnants}
\end{figure}

The stellar evolution routines included in our simulation code allow us to track star classifications throughout the run. At its final time, our simulated cluster from run \textbf{4} contains around $70,000$ white dwarfs, $700$ neutron stars and very few stellar mass black holes. The population of remnants accounts for approximately $63\%$ of the cluster mass. Figure~\ref{fig:remnants} shows the fraction of mass of white dwarfs, neutron stars, black holes, and main sequence and giant stars as a function of 3D distance from the cluster centre. The inner $0.05$ pc contained only 7 stars with a mass of $7.3 M_\odot$ in our simulation, while the inner $0.26$ pc contained about $778 M_\odot$. We therefore do not confirm the strong concentration of remnants in the very inner part of NGC 6397 predicted by MAMPOSSt-PM but do confirm the much lower density of the CMC run from~\citet{2022MNRAS.514..806V}.

Due to the advanced dynamical state of NGC~6397, the overall cluster mass is coming mainly from compact remnants. Especially the innermost 0.1~pc consist mostly of white dwarfs, which accumulate in the centre due to mass segregation.~\citet{1995ApJS..100..347D} \changetwo{compared a large set of Fokker-Planck models to observations of NGC~6397 and determined that the cluster contains $> 1400M_\odot$ in neutron stars.} Black holes and neutron stars account for very little mass \changetwo{in our model}, with the entire cluster containing only three stellar mass black holes \changetwo{and $1004M_\odot$ in neutron stars.} However this is at least in part due to the low retention fraction of 10\% against natal kicks that we assume for black holes and neutron stars in our simulations and we can't rule out slightly larger mass fractions for black holes and neutron stars.%

\begin{figure*}
    \centering
    \includegraphics[width=0.5\textwidth]{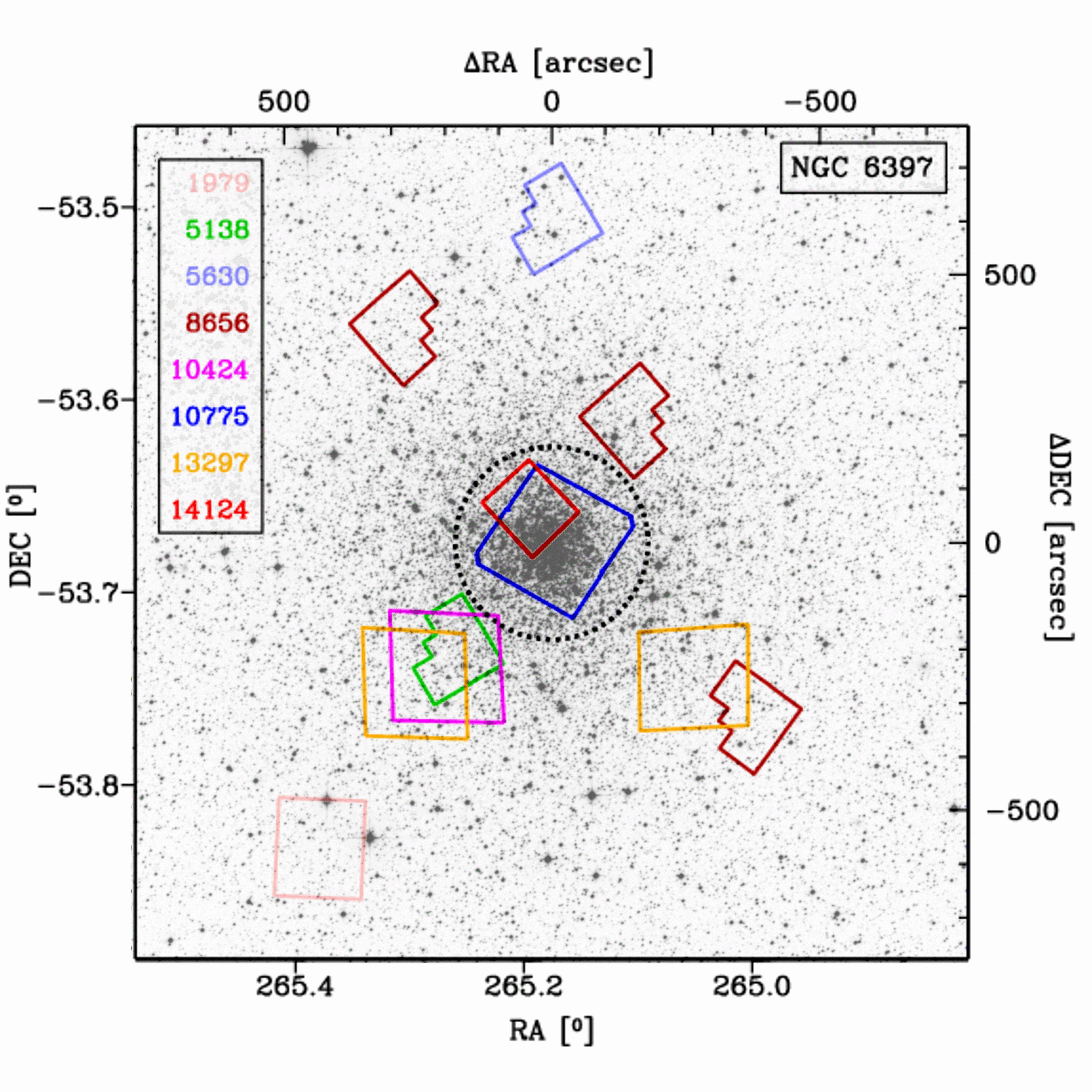}%
    \includegraphics[width=0.5\textwidth]{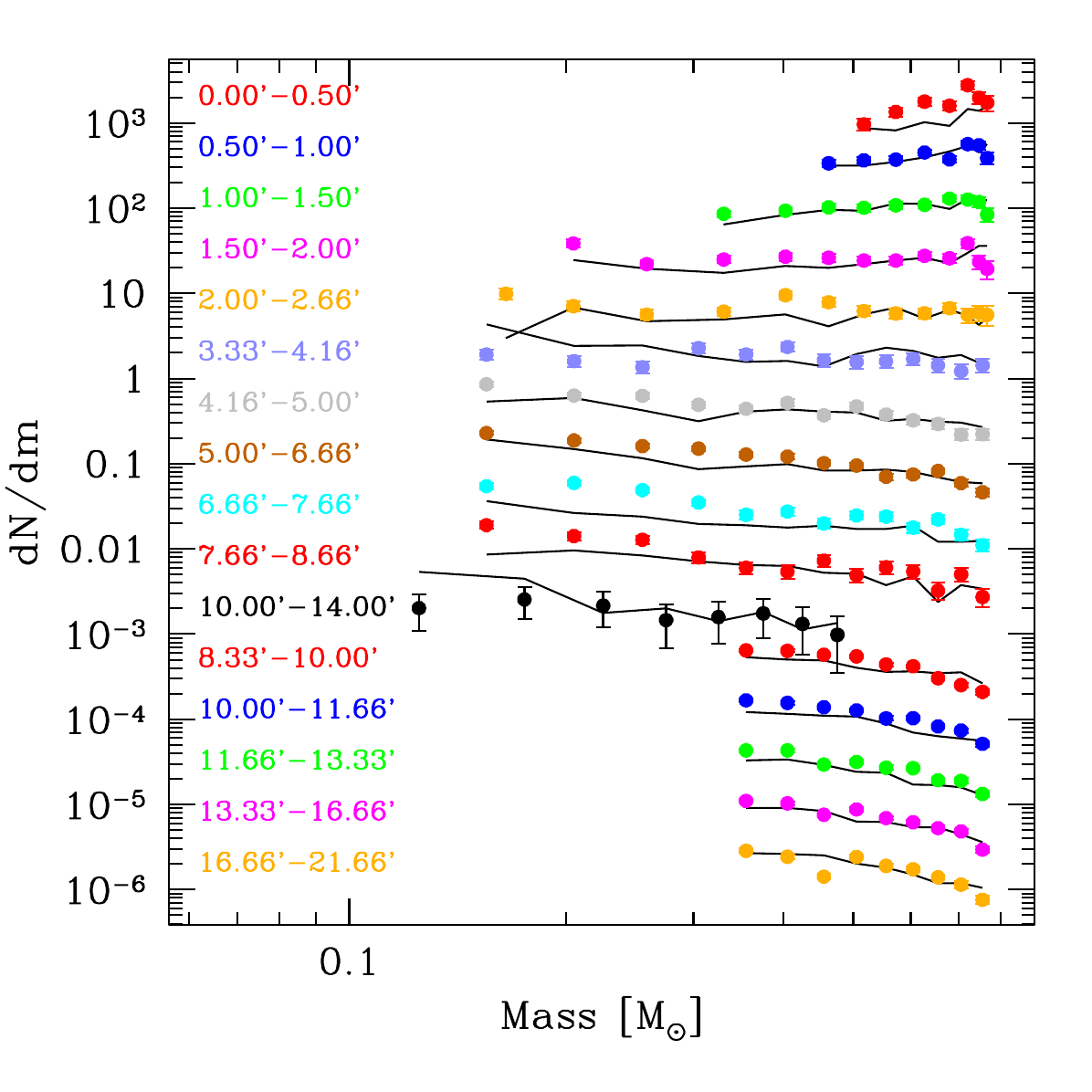}
    \caption{Left panel: Location of the HST fields in NGC 6397 for which~\citet{2023MNRAS.521.3991B} determined stellar mass functions overlaid onto a Digital Sky Survey image of the cluster. \change{The numbers indicate the HST proposal ID and the text colour corresponds to the colour of the matching field polygon(s). The dotted black circle indicates the observed half-light radius of NGC 6397.} We also show the location of the JWST field for which \citet{2024arXiv240906774L} determined the stellar mass function (proposal ID 1979). Right panel: Comparison of the stellar mass functions at different radii from~\citet{2023MNRAS.521.3991B} and \citet{2024arXiv240906774L} (circles) and the 
    simulated data from run \textbf{4} (solid lines). The $N$-body models are in good agreement with the observations both in terms of the absolute number of stars as well as their distribution across masses at the different radii.}
    \label{fig:mfunc}
\end{figure*}

In Figure~\ref{fig:mfunc} we show the stellar mass function at different radii for cluster \textbf{4} at the end of the run and compare them with observed mass function of NGC~6397 as determined by \citet{2023MNRAS.521.3991B} based on HST
data. We also add the recent JWST observations from \citet{2024arXiv240906774L} for one outer field in the cluster. \change{The observed and simulated mass function are compared directly at each radius, with no shifts to achieve alignment. Data at different radii were however shifted against each other by constant offsets to prevent all curves being directly on top of each other.} It can be seen that our model is in good agreement with the observations, both in terms of the total number of stars at different radii for which we have measurements as well as the mass distribution of the stars at different radii. It can also be seen that the distribution of stars across masses is changing from the centre towards the outer parts of NGC~6397 due to mass segregation and the simulations again match the trend seen in the observations closely. Since our simulation also fits the velocity dispersion data (Figure~\ref{fig:veldisp-nb}), which depends on total mass, and the mass function measurements constrain the total mass in visible stars, our simulation must also have the correct mass in stellar remnants. 

We hence have good agreement between observations and simulations in terms of surface density and velocity dispersion profiles as well as the stellar mass function data. Because our simulated cluster started with a standard initial mass function as described in~\citet{2023MNRAS.521.3991B}, no additional mass components like dark matter or a central IMBH are needed to explain the velocity dispersion profile of NGC 6397. We also conclude that the central dark component of NGC 6397 found by \citet{Vitral&Mamon} likely consists of stellar remnants dominated by white dwarfs, and that these were brought into the centre through mass segregation. 

\subsection{Mass loss and tidal disruption}
\label{subsec:massloss}

\begin{figure}
    \centering
    \includegraphics[width=\columnwidth]{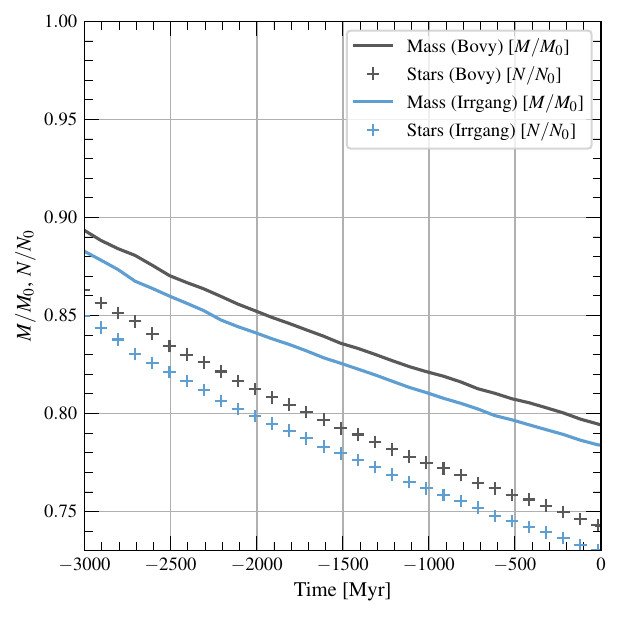}
    \caption{The ratio of bound to initial mass (stars) is shown by solid lines (crosses) for both run \textbf{4} using the ~\citet{2015ApJS..216...29B} Galactic potential and run \textbf{8} using the \citet{Irrgang2013} Galactic potential. The horizontal axis shows the time offset from the present day.}
    \label{fig:boundm}
\end{figure}

Figure~\ref{fig:boundm} depicts the final \SI{3}{\giga yr} of evolution of the bound mass for the best-fitting cluster models in the \citet{2015ApJS..216...29B} and \citet{Irrgang2013} potentials --- the initial \SI{1}{\giga yr} of evolution experienced a higher rate of mass loss due to some settling in of the cluster, and has been removed from the figure for clarity. The clusters lose over $10\%$ of their mass over the \SI{3}{\giga yr} of $N$-body simulation depicted in Fig.~\ref{fig:boundm}. Assuming a cluster age of \SI{13}{\giga yr}~\citep{2013ApJ...775..134V}, a linear decrease in mass~\citep{2003MNRAS.340..227B}, and that stellar evolution contributes only a small amount to the  mass loss in the last \SI{3}{\giga yr} depicted, this mass evolution implies a total mass loss due to two-body relaxation and interaction with the tidal field of the Milky Way alone of nearly $67\%$ up to the current age of NGC 6397. \change{The final bound mass from run \textbf{4} is $93,451 M_{\odot}$, which agrees with the finding from}~\citet{2018MNRAS.478.1520B} \change{of $90,000 \pm 10,000 M_{\odot}$ for the mass of NGC 6397.} Taking the mass loss rate and final bound mass from run \textbf{4}, and applying a further $45\%$ mass loss \change{in the phase immediately after cluster formation where most of the mass loss due to stellar evolution will occur}~\citep{2003MNRAS.340..227B}, we derive a cluster mass of $313,000 M_{\odot}$ for NGC 6397 \change{at formation}. Assuming further that the linear mass loss continues until final dissolution, we also derive a total lifetime of \SI{20}{\giga yr} for NGC 6397. 

\begin{figure*}
    \centering
    \includegraphics[width=\textwidth]{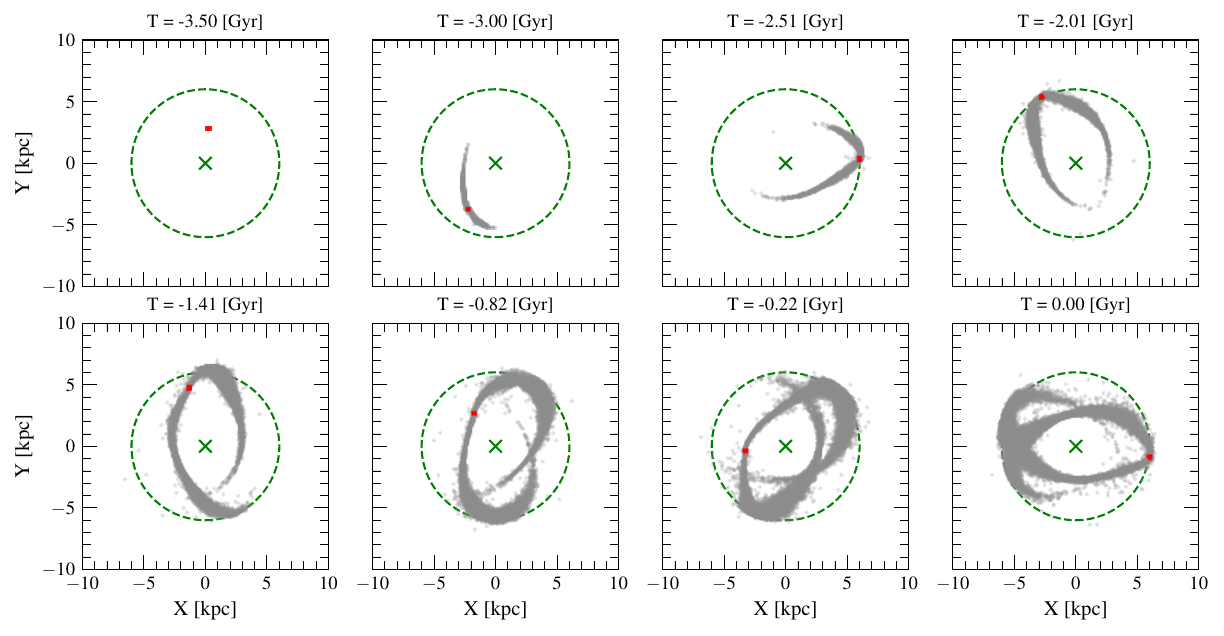}
    \caption{Eight different snapshots at different simulation times for run \textbf{4}, the best-fitting run in the ~\citet{2015ApJS..216...29B} potential. \change{Stars ejected before $T=-3.5$\,Gyr have been removed.} Bound and unbound stars are shown in red and grey, respectively. Green crosses indicate the position of the Galactic centre in each panel. The green dashed circle has a radius of $6.01$ kpc, equal to the current Galactocentric distance of NGC 6397. Since the current distance is close to the apocentre distance, the tail stars are mainly located inside this radius. A full video of run \textbf{4} is available online.}
    \label{fig:snapshot}
\end{figure*}

To visualise the tidal disruption of \change{NGC~6397} in our simulations, we show the distribution of member stars projected onto the Galactic plane for eight different times in Figure~\ref{fig:snapshot}. In this Figure one can see a rapid formation of tidal tails over \change{several} kpc within \change{1500 \,Gyr}. \change{Due to the initial settling of the cluster, as mentioned in the previous paragraph, we removed all stars that were ejected in the first $500$\,Myr after the simulation started}. Contrary to~\citet{2021MNRAS.507.1814B} who found narrow extended tidal arms within \SI{4}{\giga yr} which continued to grow, our direct simulation finds that the tidal arms quickly become widened around the orbit of NGC 6397. By the time the simulation reaches the present day, the tidal arms are poorly defined and are instead distributed all along the cluster's orbit. 

\begin{figure}
    \centering
    \includegraphics[width=\columnwidth]{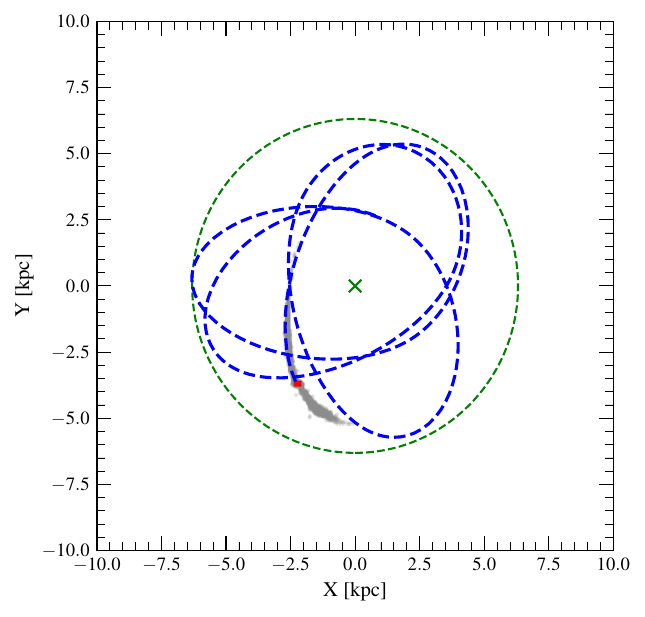}
    \caption{Tidal arms in a snapshot of run \textbf{4}, \SI{1}{\giga yr} after the start of the simulation, showing narrow tidal arms. \change{The stars shown are the same as in the top panel, second from the left in Figure}~\ref{fig:snapshot}. \change{Stars ejected before $T=-3.5$\,Gyr have been removed.} The bound and unbound stars are shown in red and grey, respectively. The green cross indicates the position of the galactic centre. The dashed blue line shows the orbit of the cluster from $T=$\SI{-3.5}{\giga yr} up until \SI{-3}{\giga yr}. The green dashed circle has a radius of $6.31$ kpc, equal to the average apogalactic distance of NGC 6397.}
    \label{fig:snapshot-past}
\end{figure}

\begin{figure}
    \centering
    \includegraphics[width=\columnwidth]{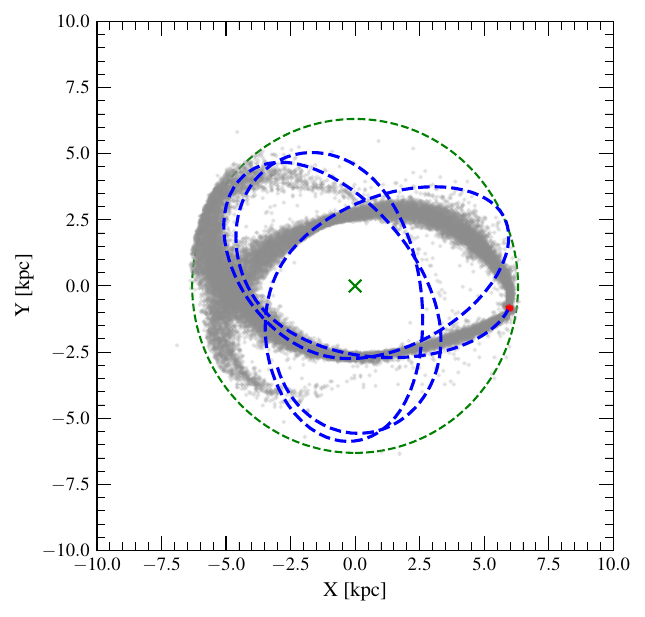}
    \caption{The same as Figure \ref{fig:snapshot-past} for the present time. \change{The stars shown are the same as in the bottom-right panel in Figure}~\ref{fig:snapshot}. The dashed blue line shows the last \SI{500}{\mega yr} of the cluster orbit. While the most recently lost stars still have a relatively narrow distribution, stars lost longer ago follow a very wide distribution.}
    \label{fig:snapshot-now}
\end{figure}

To further illustrate the early formation of tidal arms, followed by their subsequent scattering before the cluster reaches the present-day position of NGC 6397, Figures~\ref{fig:snapshot-past} and~\ref{fig:snapshot-now} show larger versions of the snapshots at \change{1\,Gyr} and \SI{4}{\giga yr} after simulation start, respectively. Additionally, these Figures show the last \SI{500}{\mega yr} of the orbital trajectory of the cluster centre.

One explanation for the widening of the distribution of stars in the tidal tails is that the stars lost are not exactly on the same orbits --- they leave the cluster at different points along the cluster's orbit and with different velocities. The top row of Figure~\ref{fig:snapshot} shows the tidal arms growing longer over time. One can see in Figure~\ref{fig:snapshot-now} that very few of the tail stars have escaped beyond the average apogalactic distance of $6.31$ kpc~\citep{2021MNRAS.505.5978V}. It is possible that, as the cluster continues along its orbit around the galactic centre, the tidal tail also continues to grow in length, leaving a wide distribution of stars all over the past and future positions of NGC 6397. We stress that, due to the perfectly static nature of the Milky Way potential used in our simulations, the diffuse tails are caused by properties of the cluster itself.%

\subsection{Near field distribution of extra-tidal stars}

\begin{figure*}
    \centering
    \includegraphics[width=\textwidth]{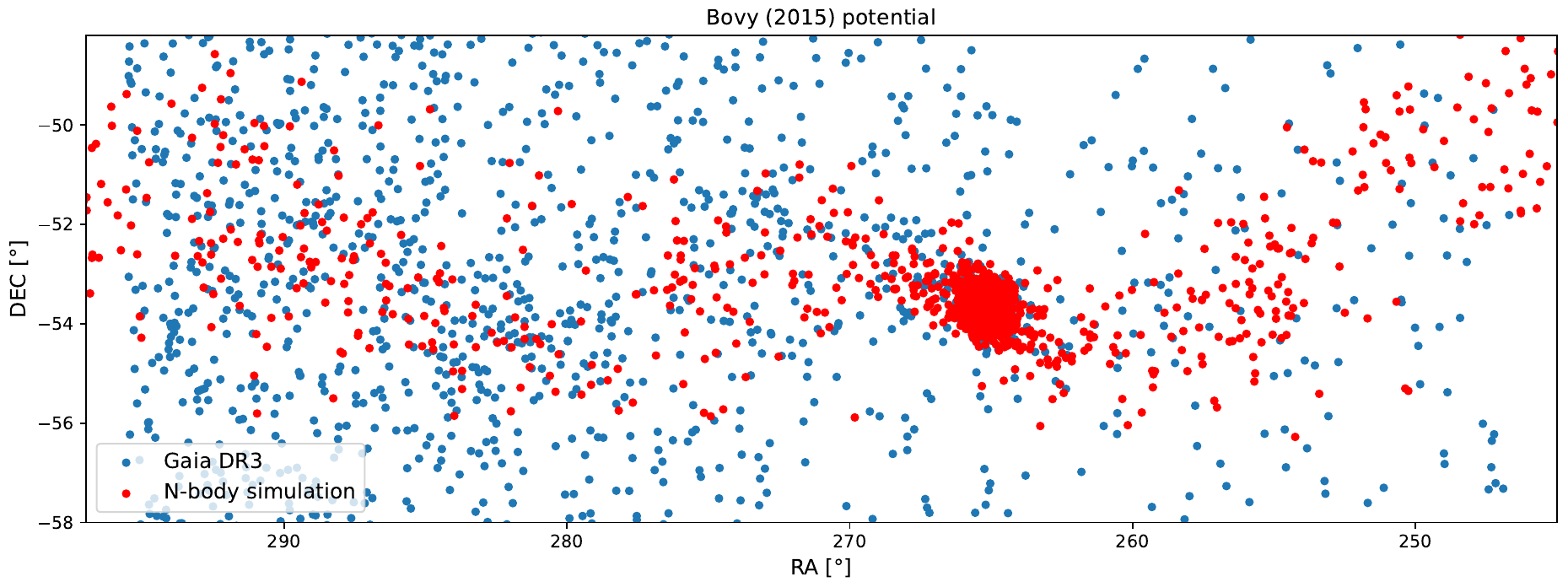}
    \includegraphics[width=\textwidth]{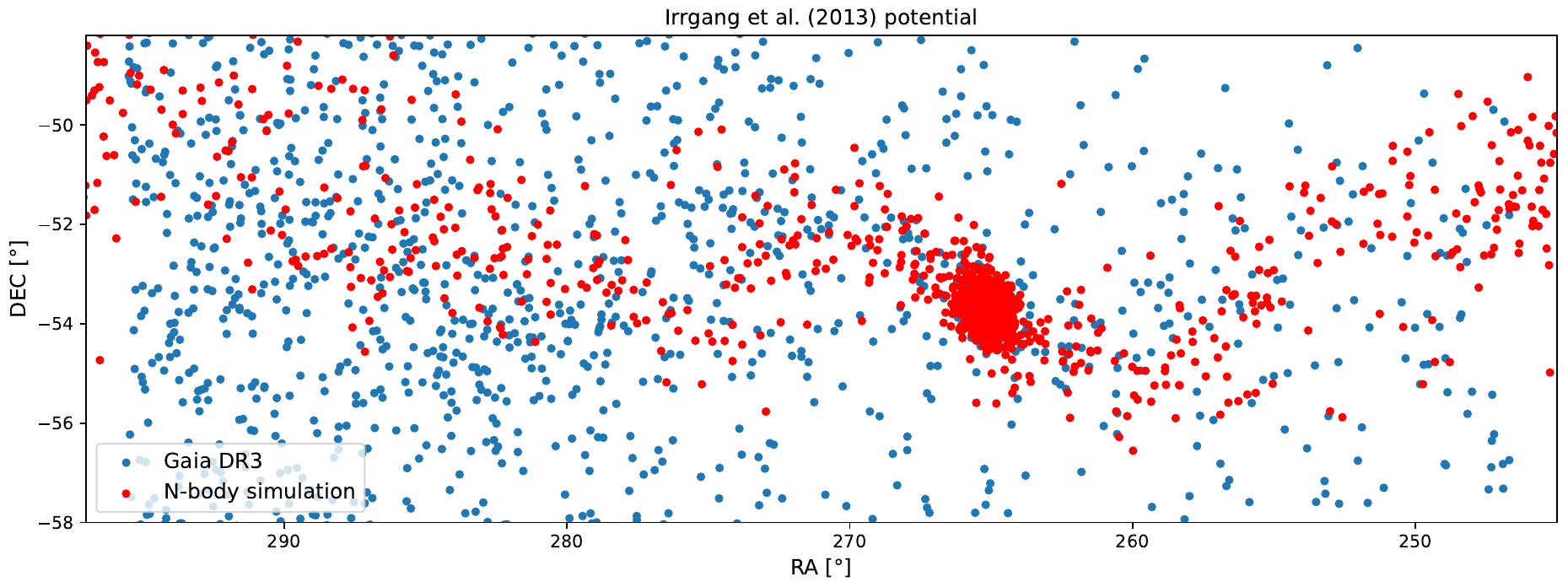}
    \caption{The location of the extra-tidal stars of NGC 6397. Stars in the simulation with the \citet{2015ApJS..216...29B} potential (run \textbf{4}) are shown in upper panel in red, those in the \citet{Irrgang2013} potential (run \textbf{8}) are shown in the lower panel in red. Possible members found in Gaia DR3 are shown in blue.}
    \label{fig:finpos}
\end{figure*}

Figure~\ref{fig:finpos} compares the location of the extra-tidal stars of NGC 6397 in the immediate vicinity of the cluster in the best-fitting $N$-body simulations with the location of possible members in Gaia DR3. For the simulated data (depicted by red circles), we show the location of all main sequence and giant stars more massive than 0.30 M$_\odot$ that are located less than 4 kpc from the Sun. The last criterion makes sure that only stars lost recently from the cluster are shown and excludes stars lost long ago that are located at much larger distances and follow different kinematics. It can be seen that the tidal tail stars in the simulations follow a wave like pattern in the sky and that there is about an equal number of stars to the left and right of the simulated cluster.

We depict the location of all possible cluster members in Gaia DR3 by blue dots in Fig.~\ref{fig:finpos}. We have selected possible members as described in Section~\ref{sec:observations} based on their proper motion, parallax and their location in a CMD. \change{Since the motion of stars in the tails is changing across the field due to perspective effects and the acceleration of stars by the Milky Way potential, we fit a low order polynomial as a function of right ascension to the proper motion of stars in the simulations and compare the predicted motion from this polynomial against the proper motion of individual Gaia DR3 stars in Equation~\eqref{eq:pmcut} to select potential members. As a result, the potential members we find in Gaia DR3 depend on the potential in which our simulations were run, explaining the slight differences in the location of the blue points in the two panels of Fig.~12.}
The possible members in Gaia DR3 are less numerous at right ascensions with RA$<260^\circ$, where the density of stars is more or less constant across the field and most stars shown are therefore likely non-members. Since these right ascensions correspond to galactic latitudes $b<5^\circ$, a possible explanation is that extinction blocks our view to the tidal tail stars of NGC~6397. On the other side of the cluster, there is an apparent overdensity of possible cluster members in the Gaia data at RA=$280^\circ$, DEC=$-55^\circ$  which seems to be connected to NGC 6397. This overdensity might also be present at larger right ascension values. The location of the overdensity is well reproduced in the run done in the 
\citet{2015ApJS..216...29B} potential, while the simulation in the
\citet{Irrgang2013} potential produces an overdensity at somewhat larger declinations than what is observed in the Gaia DR3 data. This could be an indication that it provides a less accurate description of the Galactic potential.

\change{To investigate the number of potential tidal tail stars, we took the number density of Gaia stars in the region RA$>270^\circ$ and DEC$<-56^\circ$ which is outside the tidal tail location indicated by the $N$-body simulations. We subtracted this density from the number density of stars in the region RA$>270^\circ$, DEC$>-56^\circ$, and DEC$<-49^\circ$ which is inside the tidal tail location. We found that the number of Gaia stars in the tail, after subtracting the background density, matched the number of tail stars in the simulations.}

Obtaining radial velocities for the Gaia stars shown will help to better disentangle cluster stars from Milky Way foreground stars and might also help to better constrain the Galactic potential. We finally note that~\citet{2024ApJ...967...89I} also determined the location of possible extra-tidal stars of NGC 6397 using the STREAMFINDER algorithm~\citep{10.1093/mnras/sty912}. The location of their possible extra-tidal stars agrees well with the overdensity that we find and the predicted location of these stars in the \citet{2015ApJS..216...29B} potential.%

\section{Summary and Discussion}
\label{sec:discussion}

We have investigated the state of the globular cluster NGC~6397 using Gaia DR3 data as well as dynamical $N$-body simulations. From the Gaia DR3 data we derived the density profile of NGC~6397 out to several tidal radii and found strong evidence for the presence of tidal tails around NGC~6397. We furthermore derived the velocity dispersion and surface density profiles of NGC~6397 from Gaia DR3 observations. 

We also performed a series of direct $N$-body simulations, putting simulated clusters onto the same Galactic orbit as NGC~6397 and aiming to reach a final state that resembles the real NGC~6397 as closely as possible. Our simulations contained the same number of stars as NGC~6397, avoiding the need to scale our simulations. We also included a stellar evolution component in our simulations, allowing us to accurately capture the effects of mass loss due to stellar evolution. We ran simulations assuming two different potential models for the Milky Way --- the model suggested by \citet{2015ApJS..216...29B} and the model suggested by \citet{Irrgang2013}. Compared with~\citet{2021MNRAS.507.1814B}, which did not use 1:1 simulations or stellar evolution, our simulations described the evolution of NGC 6397 more realistically which enabled us to match the current state of the cluster in terms of surface density, velocity dispersion and stellar mass function. 

Including stellar evolution also allowed us to track the classification of individual stars and investigate the population of stellar remnants in NGC 6397. We found that the inner 0.1 pc consists almost entirely of white dwarfs which accumulate in the centre due to mass segregation. Our simulations also predict that the overall mass of NGC~6397 is dominated by white dwarfs. Since our simulations started with a standard initial mass function described in~\citet{2023MNRAS.521.3991B}, we do not need additional mass components to explain the velocity dispersion profile of NGC 6397. Hence there is little room for a large mass fraction in black holes, dark matter or a central IMBH in NGC~6397. 

The best-fitting simulations of NGC 6397 ran for \SI{4}{\giga yr}, which allowed us to determine the mass loss of the cluster over long time scales. Using an initial cluster age of \SI{13}{\giga yr} given in~\citet{2013ApJ...775..134V}, we find an initial cluster mass of $3.1 \times 10^5 M_\odot$ and that NGC~6397 has lost around two thirds of its initial mass so far. We furthermore predict a total lifetime before final dissolution of \SI{20}{\giga yr} for NGC 6397. 

In our simulations, stars were ejected along the cluster's trajectory around the Milky Way, quickly leading to the formation of tidal arms which became clearly visible within \SI{500}{\mega yr} and increasingly diffuse until the cluster reached the present day location of NGC 6397. We compared the locations of the extra-tidal stars from our simulation to the possible members from Gaia DR3 and found good agreement with the locations and number of stars especially for the \citet{2015ApJS..216...29B} potential. At low right ascension values, the extra-tidal stars are hard to detect, likely due to their proximity to the galactic disc. We took into account the variations in proper motion across the field, without which the extra tidal tails do not appear. ~\citet{2021MNRAS.507.1814B} did not consider proper motion variations which, in addition to the tails being nearly impossible to detect at lower right ascension values, could explain why they did not find extra tidal stars. The good agreement between our simulations and the observations makes it unlikely that NGC 6397 is surrounded by a significant dark matter halo.  We conclude that, due to the lack of evidence for additional dark matter, the tidal tail around NGC 6397 is a result of two-body relaxation and Galactic tidal interactions.

Further observations of the tail stars of NGC 6397, including measuring their radial velocities, will help to better distinguish cluster stars from Milky Way foreground stars and may lead to stronger constraints of the Galactic potential. In addition the next Gaia data release, scheduled to be released after mid-2026, is expected to contain higher quality data which could be used to better distinguish between members and non-members and also constrain the velocity profile of stars in the tails. There is also the potential to study other globular clusters using the latest Gaia data and our advanced direct $N$-body simulation code. As shown in this work, our code is able to produce clusters in good agreement with observational data, allowing us to constrain the current state and dynamical evolution of real globular clusters.

\section*{Acknowledgements}

This work was performed on the OzSTAR national facility at Swinburne University of Technology. The OzSTAR program receives funding in part from the Astronomy National Collaborative Research Infrastructure Strategy (NCRIS) allocation provided by the Australian Government, and from the Victorian Higher Education State Investment Fund (VHESIF) provided by the Victorian Government.

\section*{Data Availability}

The data underlying this article may be made available on reasonable request to the corresponding author. Our NBODY6+P3T code is free for use and is available at \url{https://github.com/anthony-arnold/nbody6-p3t/tree/v1.2.1}.



\bibliographystyle{mnras}
\bibliography{paper} 

\begin{thebibliography}{}
\makeatletter
\relax
\def\mn@urlcharsother{\let\do\@makeother \do\$\do\&\do\#\do\^\do\_\do\%\do\~}
\def\mn@doi{\begingroup\mn@urlcharsother \@ifnextchar [ {\mn@doi@}
  {\mn@doi@[]}}
\def\mn@doi@[#1]#2{\def\@tempa{#1}\ifx\@tempa\@empty \href
  {http://dx.doi.org/#2} {doi:#2}\else \href {http://dx.doi.org/#2} {#1}\fi
  \endgroup}
\def\mn@eprint#1#2{\mn@eprint@#1:#2::\@nil}
\def\mn@eprint@arXiv#1{\href {http://arxiv.org/abs/#1} {{\tt arXiv:#1}}}
\def\mn@eprint@dblp#1{\href {http://dblp.uni-trier.de/rec/bibtex/#1.xml}
  {dblp:#1}}
\def\mn@eprint@#1:#2:#3:#4\@nil{\def\@tempa {#1}\def\@tempb {#2}\def\@tempc
  {#3}\ifx \@tempc \@empty \let \@tempc \@tempb \let \@tempb \@tempa \fi \ifx
  \@tempb \@empty \def\@tempb {arXiv}\fi \@ifundefined
  {mn@eprint@\@tempb}{\@tempb:\@tempc}{\expandafter \expandafter \csname
  mn@eprint@\@tempb\endcsname \expandafter{\@tempc}}}

\bibitem[\protect\citeauthoryear{{Arnold}, {Baumgardt}  \& {Wang}}{{Arnold}
  et~al.}{2022}]{2022MNRAS.509.2075A}
{Arnold} A.~D.,  {Baumgardt} H.,   {Wang} L.,  2022, \mn@doi [\mnras]
  {10.1093/mnras/stab3090}, \href
  {https://ui.adsabs.harvard.edu/abs/2022MNRAS.509.2075A} {509, 2075}

\bibitem[\protect\citeauthoryear{{Baumgardt} \& {Hilker}}{{Baumgardt} \&
  {Hilker}}{2018}]{2018MNRAS.478.1520B}
{Baumgardt} H.,  {Hilker} M.,  2018, \mn@doi [\mnras] {10.1093/mnras/sty1057},
  \href {https://ui.adsabs.harvard.edu/abs/2018MNRAS.478.1520B} {478, 1520}

\bibitem[\protect\citeauthoryear{{Baumgardt} \& {Makino}}{{Baumgardt} \&
  {Makino}}{2003}]{2003MNRAS.340..227B}
{Baumgardt} H.,  {Makino} J.,  2003, \mn@doi [\mnras]
  {10.1046/j.1365-8711.2003.06286.x}, \href
  {https://ui.adsabs.harvard.edu/abs/2003MNRAS.340..227B} {340, 227}

\bibitem[\protect\citeauthoryear{{Baumgardt} \& {Vasiliev}}{{Baumgardt} \&
  {Vasiliev}}{2021}]{2021MNRAS.505.5957B}
{Baumgardt} H.,  {Vasiliev} E.,  2021, \mn@doi [\mnras]
  {10.1093/mnras/stab1474}, \href
  {https://ui.adsabs.harvard.edu/abs/2021MNRAS.505.5957B} {505, 5957}

\bibitem[\protect\citeauthoryear{{Baumgardt}, {Hilker}, {Sollima}  \&
  {Bellini}}{{Baumgardt} et~al.}{2019}]{2019MNRAS.482.5138B}
{Baumgardt} H.,  {Hilker} M.,  {Sollima} A.,   {Bellini} A.,  2019, \mn@doi
  [\mnras] {10.1093/mnras/sty2997}, \href
  {https://ui.adsabs.harvard.edu/abs/2019MNRAS.482.5138B} {482, 5138}

\bibitem[\protect\citeauthoryear{{Baumgardt}, {H{\'e}nault-Brunet}, {Dickson}
  \& {Sollima}}{{Baumgardt} et~al.}{2023}]{2023MNRAS.521.3991B}
{Baumgardt} H.,  {H{\'e}nault-Brunet} V.,  {Dickson} N.,   {Sollima} A.,  2023,
  \mn@doi [\mnras] {10.1093/mnras/stad631}, \href
  {https://ui.adsabs.harvard.edu/abs/2023MNRAS.521.3991B} {521, 3991}

\bibitem[\protect\citeauthoryear{{B{\'e}dorf}, {Gaburov}  \& {Portegies
  Zwart}}{{B{\'e}dorf} et~al.}{2012}]{2012ASPC..453..325B}
{B{\'e}dorf} J.,  {Gaburov} E.,   {Portegies Zwart} S.,  2012, in
  {Capuzzo-Dolcetta} R.,  {Limongi} M.,   {Tornamb{\`e}} A.,  eds,
  Astronomical Society of the Pacific Conference Series Vol. 453, Advances in
  Computational Astrophysics: Methods, Tools, and Outcome. p.~325 (\mn@eprint
  {arXiv} {1204.2280}), \mn@doi{10.48550/arXiv.1204.2280}

\bibitem[\protect\citeauthoryear{{Boldrini} \& {Vitral}}{{Boldrini} \&
  {Vitral}}{2021}]{2021MNRAS.507.1814B}
{Boldrini} P.,  {Vitral} E.,  2021, \mn@doi [\mnras] {10.1093/mnras/stab2035},
  \href {https://ui.adsabs.harvard.edu/abs/2021MNRAS.507.1814B} {507, 1814}

\bibitem[\protect\citeauthoryear{{Bovy}}{{Bovy}}{2015}]{2015ApJS..216...29B}
{Bovy} J.,  2015, \mn@doi [\apjs] {10.1088/0067-0049/216/2/29}, \href
  {https://ui.adsabs.harvard.edu/abs/2015ApJS..216...29B} {216, 29}

\bibitem[\protect\citeauthoryear{{Bressan}, {Marigo}, {Girardi}, {Salasnich},
  {Dal Cero}, {Rubele}  \& {Nanni}}{{Bressan}
  et~al.}{2012}]{2012MNRAS.427..127B}
{Bressan} A.,  {Marigo} P.,  {Girardi} L.,  {Salasnich} B.,  {Dal Cero} C.,
  {Rubele} S.,   {Nanni} A.,  2012, \mn@doi [\mnras]
  {10.1111/j.1365-2966.2012.21948.x}, \href
  {https://ui.adsabs.harvard.edu/abs/2012MNRAS.427..127B} {427, 127}

\bibitem[\protect\citeauthoryear{{Cantat-Gaudin} et~al.,}{{Cantat-Gaudin}
  et~al.}{2023}]{2023A&A...669A..55C}
{Cantat-Gaudin} T.,  et~al., 2023, \mn@doi [\aap]
  {10.1051/0004-6361/202244784}, \href
  {https://ui.adsabs.harvard.edu/abs/2023A&A...669A..55C} {669, A55}

\bibitem[\protect\citeauthoryear{Carlberg \& Grillmair}{Carlberg \&
  Grillmair}{2021}]{Carlberg_2021}
Carlberg R.~G.,  Grillmair C.~J.,  2021, \mn@doi [\apj]
  {10.3847/1538-4357/ac289f}, 922, 104

\bibitem[\protect\citeauthoryear{Carlberg \& Grillmair}{Carlberg \&
  Grillmair}{2022}]{Carlberg_2022}
Carlberg R.~G.,  Grillmair C.~J.,  2022, \mn@doi [\apj]
  {10.3847/1538-4357/ac7d54}, 935, 14

\bibitem[\protect\citeauthoryear{Chomiuk, Strader, Maccarone, Miller-Jones,
  Heinke, Noyola, Seth  \& Ransom}{Chomiuk et~al.}{2013}]{Chomiuk_2013}
Chomiuk L.,  Strader J.,  Maccarone T.~J.,  Miller-Jones J. C.~A.,  Heinke C.,
  Noyola E.,  Seth A.~C.,   Ransom S.,  2013, \mn@doi [\apj]
  {10.1088/0004-637X/777/1/69}, 777, 69

\bibitem[\protect\citeauthoryear{{Drukier}}{{Drukier}}{1995}]{1995ApJS..100..347D}
{Drukier} G.~A.,  1995, \mn@doi [\apjs] {10.1086/192223}, \href
  {https://ui.adsabs.harvard.edu/abs/1995ApJS..100..347D} {100, 347}

\bibitem[\protect\citeauthoryear{Freire}{Freire}{2012}]{2013IAUS..291..243F}
Freire P. C.~C.,  2012, \mn@doi [Proceedings of the International Astronomical
  Union] {10.1017/S1743921312023770}, 8, 243–250

\bibitem[\protect\citeauthoryear{{Gaia Collaboration} et~al.,}{{Gaia
  Collaboration} et~al.}{2023}]{2023A&A...674A...1G}
{Gaia Collaboration} et~al., 2023, \mn@doi [\aap]
  {10.1051/0004-6361/202243940}, \href
  {https://ui.adsabs.harvard.edu/abs/2023A&A...674A...1G} {674, A1}

\bibitem[\protect\citeauthoryear{{Giesers} et~al.,}{{Giesers}
  et~al.}{2018}]{2018MNRAS.475L..15G}
{Giesers} B.,  et~al., 2018, \mn@doi [\mnras] {10.1093/mnrasl/slx203}, \href
  {https://ui.adsabs.harvard.edu/abs/2018MNRAS.475L..15G} {475, L15}

\bibitem[\protect\citeauthoryear{{Goldsbury}, {Richer}, {Anderson}, {Dotter},
  {Sarajedini}  \& {Woodley}}{{Goldsbury} et~al.}{2010}]{2010AJ....140.1830G}
{Goldsbury} R.,  {Richer} H.~B.,  {Anderson} J.,  {Dotter} A.,  {Sarajedini}
  A.,   {Woodley} K.,  2010, \mn@doi [\aj] {10.1088/0004-6256/140/6/1830},
  \href {https://ui.adsabs.harvard.edu/abs/2010AJ....140.1830G} {140, 1830}

\bibitem[\protect\citeauthoryear{H{\"a}berle et~al.,}{H{\"a}berle
  et~al.}{2024}]{Häberle2024}
H{\"a}berle M.,  et~al., 2024, \mn@doi [Nature] {10.1038/s41586-024-07511-z},
  631, 285

\bibitem[\protect\citeauthoryear{{Heggie} \& {Giersz}}{{Heggie} \&
  {Giersz}}{2009}]{2009MNRAS.397L..46H}
{Heggie} D.~C.,  {Giersz} M.,  2009, \mn@doi [\mnras]
  {10.1111/j.1745-3933.2009.00681.x}, \href
  {https://ui.adsabs.harvard.edu/abs/2009MNRAS.397L..46H} {397, L46}

\bibitem[\protect\citeauthoryear{{Heggie} \& {Hut}}{{Heggie} \&
  {Hut}}{1996}]{1996IAUS..174..303H}
{Heggie} D.~C.,  {Hut} P.,  1996, in {Hut} P.,  {Makino} J.,  eds,  IAU
  Symposium Vol. 174, Dynamical Evolution of Star Clusters: Confrontation of
  Theory and Observations. p.~303 (\mn@eprint {arXiv} {astro-ph/9511115}),
  \mn@doi{10.48550/arXiv.astro-ph/9511115}

\bibitem[\protect\citeauthoryear{{Hurley}, {Pols}  \& {Tout}}{{Hurley}
  et~al.}{2000}]{2000MNRAS.315..543H}
{Hurley} J.~R.,  {Pols} O.~R.,   {Tout} C.~A.,  2000, \mn@doi [\mnras]
  {10.1046/j.1365-8711.2000.03426.x}, \href
  {https://ui.adsabs.harvard.edu/abs/2000MNRAS.315..543H} {315, 543}

\bibitem[\protect\citeauthoryear{{Hurley}, {Tout}  \& {Pols}}{{Hurley}
  et~al.}{2002}]{2002MNRAS.329..897H}
{Hurley} J.~R.,  {Tout} C.~A.,   {Pols} O.~R.,  2002, \mn@doi [\mnras]
  {10.1046/j.1365-8711.2002.05038.x}, \href
  {https://ui.adsabs.harvard.edu/abs/2002MNRAS.329..897H} {329, 897}

\bibitem[\protect\citeauthoryear{{Ibata} et~al.,}{{Ibata}
  et~al.}{2021}]{2021ApJ...914..123I}
{Ibata} R.,  et~al., 2021, \mn@doi [\apj] {10.3847/1538-4357/abfcc2}, \href
  {https://ui.adsabs.harvard.edu/abs/2021ApJ...914..123I} {914, 123}

\bibitem[\protect\citeauthoryear{{Ibata} et~al.,}{{Ibata}
  et~al.}{2024}]{2024ApJ...967...89I}
{Ibata} R.,  et~al., 2024, \mn@doi [\apj] {10.3847/1538-4357/ad382d}, \href
  {https://ui.adsabs.harvard.edu/abs/2024ApJ...967...89I} {967, 89}

\bibitem[\protect\citeauthoryear{{Irrgang}, {Wilcox}, {Tucker}  \&
  {Schiefelbein}}{{Irrgang} et~al.}{2013}]{Irrgang2013}
{Irrgang} A.,  {Wilcox} B.,  {Tucker} E.,   {Schiefelbein} L.,  2013, \mn@doi
  [\aap] {10.1051/0004-6361/201220540}, \href
  {https://ui.adsabs.harvard.edu/abs/2013A&A...549A.137I} {549, A137}

\bibitem[\protect\citeauthoryear{{Kamann} et~al.,}{{Kamann}
  et~al.}{2016}]{2016A&A...588A.149K}
{Kamann} S.,  et~al., 2016, \mn@doi [\aap] {10.1051/0004-6361/201527065}, \href
  {https://ui.adsabs.harvard.edu/abs/2016A&A...588A.149K} {588, A149}

\bibitem[\protect\citeauthoryear{{Libralato} et~al.,}{{Libralato}
  et~al.}{2022}]{2022ApJ...934..150L}
{Libralato} M.,  et~al., 2022, \mn@doi [\apj] {10.3847/1538-4357/ac7727}, \href
  {https://ui.adsabs.harvard.edu/abs/2022ApJ...934..150L} {934, 150}

\bibitem[\protect\citeauthoryear{{Libralato} et~al.,}{{Libralato}
  et~al.}{2024}]{2024arXiv240906774L}
{Libralato} M.,  et~al., 2024, \mn@doi [arXiv e-prints]
  {10.48550/arXiv.2409.06774}, \href
  {https://ui.adsabs.harvard.edu/abs/2024arXiv240906774L} {p. arXiv:2409.06774}

\bibitem[\protect\citeauthoryear{Malhan \& Ibata}{Malhan \&
  Ibata}{2018}]{10.1093/mnras/sty912}
Malhan K.,  Ibata R.~A.,  2018, \mn@doi [\mnras] {10.1093/mnras/sty912}, 477,
  4063

\bibitem[\protect\citeauthoryear{{Mamon}, {Biviano}  \& {Bou{\'e}}}{{Mamon}
  et~al.}{2013}]{2013MNRAS.429.3079M}
{Mamon} G.~A.,  {Biviano} A.,   {Bou{\'e}} G.,  2013, \mn@doi [\mnras]
  {10.1093/mnras/sts565}, \href
  {https://ui.adsabs.harvard.edu/abs/2013MNRAS.429.3079M} {429, 3079}

\bibitem[\protect\citeauthoryear{Miki \& Umemura}{Miki \&
  Umemura}{2017}]{Miki_2017}
Miki Y.,  Umemura M.,  2017, \mn@doi [New Astronomy]
  {10.1016/j.newast.2016.10.007}, 52, 65–81

\bibitem[\protect\citeauthoryear{Nitadori \& Aarseth}{Nitadori \&
  Aarseth}{2012}]{10.1111/j.1365-2966.2012.21227.x}
Nitadori K.,  Aarseth S.~J.,  2012, \mn@doi [\mnras]
  {10.1111/j.1365-2966.2012.21227.x}, 424, 545

\bibitem[\protect\citeauthoryear{{Pe{\~n}arrubia}, {Varri}, {Breen}, {Ferguson}
   \& {S{\'a}nchez-Janssen}}{{Pe{\~n}arrubia}
  et~al.}{2017}]{2017MNRAS.471L..31P}
{Pe{\~n}arrubia} J.,  {Varri} A.~L.,  {Breen} P.~G.,  {Ferguson} A. M.~N.,
  {S{\'a}nchez-Janssen} R.,  2017, \mn@doi [\mnras] {10.1093/mnrasl/slx094},
  \href {https://ui.adsabs.harvard.edu/abs/2017MNRAS.471L..31P} {471, L31}

\bibitem[\protect\citeauthoryear{Ransom}{Ransom}{2007}]{2008IAUS..246..291R}
Ransom S.~M.,  2007, \mn@doi [Proceedings of the International Astronomical
  Union] {10.1017/S1743921308015810}, 3, 291–300

\bibitem[\protect\citeauthoryear{{Rodriguez} et~al.,}{{Rodriguez}
  et~al.}{2022}]{2022ApJS..258...22R}
{Rodriguez} C.~L.,  et~al., 2022, \mn@doi [\apjs] {10.3847/1538-4365/ac2edf},
  \href {https://ui.adsabs.harvard.edu/abs/2022ApJS..258...22R} {258, 22}

\bibitem[\protect\citeauthoryear{{Sarajedini} et~al.,}{{Sarajedini}
  et~al.}{2007}]{2007AJ....133.1658S}
{Sarajedini} A.,  et~al., 2007, \mn@doi [\aj] {10.1086/511979}, \href
  {https://ui.adsabs.harvard.edu/abs/2007AJ....133.1658S} {133, 1658}

\bibitem[\protect\citeauthoryear{{Stetson}, {Pancino}, {Zocchi}, {Sanna}  \&
  {Monelli}}{{Stetson} et~al.}{2019}]{2019MNRAS.485.3042S}
{Stetson} P.~B.,  {Pancino} E.,  {Zocchi} A.,  {Sanna} N.,   {Monelli} M.,
  2019, \mn@doi [\mnras] {10.1093/mnras/stz585}, \href
  {https://ui.adsabs.harvard.edu/abs/2019MNRAS.485.3042S} {485, 3042}

\bibitem[\protect\citeauthoryear{{VandenBerg}, {Brogaard}, {Leaman}  \&
  {Casagrande}}{{VandenBerg} et~al.}{2013}]{2013ApJ...775..134V}
{VandenBerg} D.~A.,  {Brogaard} K.,  {Leaman} R.,   {Casagrande} L.,  2013,
  \mn@doi [\apj] {10.1088/0004-637X/775/2/134}, \href
  {https://ui.adsabs.harvard.edu/abs/2013ApJ...775..134V} {775, 134}

\bibitem[\protect\citeauthoryear{{Vasiliev} \& {Baumgardt}}{{Vasiliev} \&
  {Baumgardt}}{2021}]{2021MNRAS.505.5978V}
{Vasiliev} E.,  {Baumgardt} H.,  2021, \mn@doi [\mnras]
  {10.1093/mnras/stab1475}, \href
  {https://ui.adsabs.harvard.edu/abs/2021MNRAS.505.5978V} {505, 5978}

\bibitem[\protect\citeauthoryear{{Vitral} \& {Mamon}}{{Vitral} \&
  {Mamon}}{2021}]{Vitral&Mamon}
{Vitral} E.,  {Mamon} G.~A.,  2021, \mn@doi [\aap]
  {10.1051/0004-6361/202039650}, \href
  {https://ui.adsabs.harvard.edu/abs/2021A&A...646A..63V} {646, A63}

\bibitem[\protect\citeauthoryear{{Vitral}, {Kremer}, {Libralato}, {Mamon}  \&
  {Bellini}}{{Vitral} et~al.}{2022}]{2022MNRAS.514..806V}
{Vitral} E.,  {Kremer} K.,  {Libralato} M.,  {Mamon} G.~A.,   {Bellini} A.,
  2022, \mn@doi [\mnras] {10.1093/mnras/stac1337}, \href
  {https://ui.adsabs.harvard.edu/abs/2022MNRAS.514..806V} {514, 806}

\bibitem[\protect\citeauthoryear{Wan et~al.,}{Wan
  et~al.}{2021}]{10.1093/mnras/stab306}
Wan Z.,  et~al., 2021, \mn@doi [\mnras] {10.1093/mnras/stab306}, 502, 4513

\bibitem[\protect\citeauthoryear{{Wan} et~al.,}{{Wan}
  et~al.}{2023}]{2023MNRAS.519..192W}
{Wan} Z.,  et~al., 2023, \mn@doi [\mnras] {10.1093/mnras/stac3566}, \href
  {https://ui.adsabs.harvard.edu/abs/2023MNRAS.519..192W} {519, 192}

\bibitem[\protect\citeauthoryear{{de Boer}, {Gieles}, {Balbinot},
  {H{\'e}nault-Brunet}, {Sollima}, {Watkins}  \& {Claydon}}{{de Boer}
  et~al.}{2019}]{2019MNRAS.485.4906D}
{de Boer} T.~J.~L.,  {Gieles} M.,  {Balbinot} E.,  {H{\'e}nault-Brunet} V.,
  {Sollima} A.,  {Watkins} L.~L.,   {Claydon} I.,  2019, \mn@doi [\mnras]
  {10.1093/mnras/stz651}, \href
  {https://ui.adsabs.harvard.edu/abs/2019MNRAS.485.4906D} {485, 4906}

\makeatother
\end{thebibliography}




\appendix

\section{Surface Density Profile}

Table~\ref{tab:sb} shows the data from Figure~\ref{fig:sb} up to $r=\ang{12.5}$. For brevity, not all data points are shown - the full table is available at \url{https://github.com/anthony-arnold/nbody6-p3t/blob/v1.2.1/profiles/ngc6397.txt}.

\begin{table}
\caption{The background subtracted surface density profile of NGC 6397 \change{described in Section}~\ref{subsec:sb}. From left to right the columns contain: the radius; \change{the number of stars in the radial bin}; \changetwo{the background subtracted surface density in units of the number of stars per square arcsec}; the surface density error; the source of the data. The source numbers in the final column correspond to: 1 =~\citet{2007AJ....133.1658S}; 2 =~\citet{2019MNRAS.485.3042S}; 3 = this work.}
\label{tab:sb}
\begin{tabularx}{\columnwidth}{ccccc} 
\hline
\thead{$\log r$ \arcsec} & 
\thead{$N$} &
\thead{\changetwo{SD [$N $\arcsec$^{-2}$]}} & 
\thead{\changetwo{$\sigma$ SD}} & 
\thead{Data Source}\\
\hline
0.7 & 74 & 7.45e-01 & 8.66e-02 & 1\\
0.8 & 31 & 5.33e-01 & 9.58e-02 & 1\\
0.9 & 42 & 4.56e-01 & 7.04e-02 & 1\\
1.0 & 35 & 2.40e-01 & 4.05e-02 & 1\\
1.1 & 79 & 3.42e-01 & 3.84e-02 & 1\\
1.2 & 89 & 2.43e-01 & 2.57e-02 & 1\\
1.3 & 106 & 1.82e-01 & 1.77e-02 & 1\\
1.4 & 141 & 1.53e-01 & 1.29e-02 & 1\\
1.5 & 213 & 1.46e-01 & 1.00e-02 & 1\\
1.6 & 254 & 1.10e-01 & 6.89e-03 & 1\\
1.7 & 344 & 9.38e-02 & 5.06e-03 & 1\\
1.8 & 376 & 6.47e-02 & 3.34e-03 & 1\\
1.9 & 526 & 5.71e-02 & 2.49e-03 & 1\\
2.0 & 565 & 3.87e-02 & 1.63e-03 & 1\\
2.1 & 542 & 2.34e-02 & 1.01e-03 & 2\\
2.2 & 597 & 1.63e-02 & 6.66e-04 & 2\\
2.3 & 730 & 1.26e-02 & 4.65e-04 & 2\\
2.4 & 821 & 8.91e-03 & 3.11e-04 & 2\\
2.5 & 887 & 6.08e-03 & 2.04e-04 & 2\\
2.6 & 938 & 4.06e-03 & 1.32e-04 & 2\\
2.7 & 638 & 1.74e-03 & 6.89e-05 & 3\\
2.8 & 602 & 1.04e-03 & 4.22e-05 & 3\\
2.9 & 560 & 6.08e-04 & 2.57e-05 & 3\\
3.0 & 372 & 2.54e-04 & 1.32e-05 & 3\\
3.1 & 236 & 1.01e-04 & 6.64e-06 & 3\\
3.2 & 144 & 3.86e-05 & 3.27e-06 & 3\\
3.3 & 41 & 6.41e-06 & 1.10e-06 & 3\\
3.4 & 17 & 1.21e-06 & 4.48e-07 & 3\\
3.5 & 14 & 3.19e-07 & 2.56e-07 & 3\\

\hline
\end{tabularx}
\end{table}



\bsp	
\label{lastpage}
\end{document}